\documentclass[12pt]{article}
\usepackage{geometry}
\usepackage{makeidx}
\usepackage[T1]{fontenc}
\usepackage[dvipsnames,svgnames,table]{xcolor}
\usepackage{epstopdf}
\usepackage{epsfig}
\usepackage{textcomp}
\usepackage{hyperref}
\usepackage{amsmath}
\usepackage{amssymb}
\usepackage{multirow}
\usepackage{multicol}
\usepackage{booktabs}
\usepackage{lastpage}
\usepackage{hyperref} 
\usepackage{graphicx}
\usepackage{enumerate}
\usepackage{threeparttable}
\usepackage{float}
\usepackage{array}
\usepackage{cite}
\usepackage{color,soul}
\usepackage{listings}
\usepackage{tikz}
\usepackage{longtable}
\makeatletter
\renewcommand\section{\@startsection{section}{1}{\z@}%
{-2.5ex \@plus -1ex \@minus -.2ex}%
{2.3ex \@plus.2ex}%
{\normalfont\large\bfseries}}
\renewcommand\subsection{\@startsection{subsection}{1}{\z@}%
{-2.5ex \@plus -1ex \@minus -.2ex}%
{2.3ex \@plus.2ex}%
{\small\bfseries}}

\begin{document}
\bigskip
\title{\textbf{Thermal/epithermal neutron detection via LiO$_{2}$-doped fiber bundles: A computational study through GEANT4 simulations}}
\medskip
\author{\small A. Ilker Topuz$^{1}$}
\medskip
\date{\small$^1$Department of Physics, SRM University-AP, Amaravati 522240, India\\email: ahmetllker.t@srmap.in, aitopuz@zohomail.eu}
\maketitle
\begin{abstract}
One of the neutron capture processes in lithium or lithium-doped materials leads to the production of secondary alpha-triton pairs. The counts of these secondary charged particle pairs in the presence of thermal/epithermal neutrons indirectly give access to detecting the primary incident neutrons with the aid of the plastic scintillators by reminding that both these secondary alphas and tritons deposit their entire kinetic energy within a short range. In this study, we modify the polystyrene fiber core of Kuraray Y11-200(M) by doping with 10 wt.$\%$ LiO$_{2}$. We investigate a fiber bundle system hinged on 10 wt.$\%$ LiO$_{2}$-doped Kuraray Y11-200(M) fibers via the GEANT4 simulations by targeting the detection of the incident thermal/epithermal neutrons through the generation of the secondary alpha-triton pairs. First, we estimate an optimal number of fiber arrays by using a monoenergetic beam of thermal neutrons with a kinetic energy of 0.025 eV and we count the number of secondary alpha-triton pairs. Subsequently, we insert a pixelated photo-sensor on both ends of the fiber bundle in order to collect the entrapped optical photons. In addition to the thermal neutrons, we also explore the detection of epithermal neutrons with a kinetic energy of 1 eV by utilizing a combination of the 10 wt.$\%$ LiO$_{2}$-doped Kuraray Y11-200(M) fiber bundles and the neutron absorption layers. We demonstrate that the LiO$_{2}$-doped plastic scintillating fiber bundles are instrumental for the detection of thermal/epithermal neutrons in diverse applications.
\end{abstract}
\textbf{\textit{Keywords: }} Thermal neutrons; Epithermal neutrons; Secondary particles; Lithium oxide; Plastic scintillators; Pixel detectors; GEANT4.
\section{Introduction}
\label{Intro}
The detection of thermal/epithermal neutrons is a fundamental aspect for the nuclear applications including but not limited to space dosimetry, activation analysis, and reactor instrumentation. However, the neutron detection systems axiomatically rely on a conversion process where an interaction of the incident neutrons with a nucleus stochastically results in the generation of a secondary particle~\cite{peurrung2000recent, pietropaolo2020neutron}. For this purpose, among the prevalent materials used to provoke the neutron interactions is lithium~\cite{mayer2015development,rich2015fabrication}, where one of the thermal/epithermal neutron capture processes gives rise to the production of the secondary alpha-triton pairs as described in
\begin{equation}
\mbox{$^{6}$Li + n $\longrightarrow$ $^{4}$He(2.05~MeV) + $^{3}$H(2.73~MeV)}
\label{lithiumneutronabsorption}
\end{equation}
By recalling that the secondary charged particles generated in Eq.~(\ref{lithiumneutronabsorption}) lose their entire kinetic energies within a few micrometers~\cite{ziegler2010srim, mayer2019optimization}, the plastic scintillators that contain a certain amount of lithium release a significant number of optical photons due to the complete energy deposition of these secondary charged particles, thereby indicating the presence of the thermal/epithermal neutrons. 

The introduction of lithium brings several advantages regarding the neutron detection. In the first instance, lithium has a high cross-section for the neutron capture, particularly for the thermal neutrons. Additionally, the released secondary alpha particles and tritons possess a high linear energy transfer by ensuring that the energy is deposited within a limited volume, which is crucial for the accurate detection. The advantages of lithium are combined with the mechanical flexibility and durability of fiber-based detectors when lithium is added to the plastic scintillating fibers. The scintillating fibers retain a high surface-area-to-volume ratio that increases the likelihood of interacting with neutrons. Additionally, these fibers can be easily manufactured into different shapes and sizes by providing flexible detector designs that can fit into various applications and environments.

In this study, we explore a fiber bundle system built on the Kuraray Y11-200(M) fibers~\cite{Kuraray} that are additionally doped with 10 wt.$\%$ LiO$_{2}$ via the GEANT4 simulations~\cite{agostinelli2003geant4, TopuzGithubLiO2doped} by aiming at detecting the incident thermal/epithermal neutrons through the emission of the secondary alpha-triton pairs. First, we approximate an ideal number of fiber arrays by using a monoenergetic beam of thermal neutrons with a kinetic energy of 0.025 eV and we track the secondary alpha-triton pairs. Second, we introduce a pixelated photo-detector on each side of the fiber bundle in order to record the optical photons. Along with the thermal neutrons, we also investigate the identification of the epithermal neutrons by utilizing a combination of the 10 wt.$\%$ LiO$_{2}$-doped Kuraray Y11-200(M) fiber bundles and the neutron absorption layers. This study is organized as follows. In section~\ref{Setup}, we state the features of the 10 wt.$\%$ LiO$_{2}$-doped Kuraray Y11-200(M) fiber along with the simulation properties in GEANT4, while the simulation outcomes are exhibited in section~\ref{Outcomes}. Finally, section~\ref{Conclusion} incorporates our conclusions drawn from our GEANT4 simulations.
\section{Simulation setup}
\label{Setup}
\begin{figure}[H]
\begin{center}
\includegraphics[width=12cm]{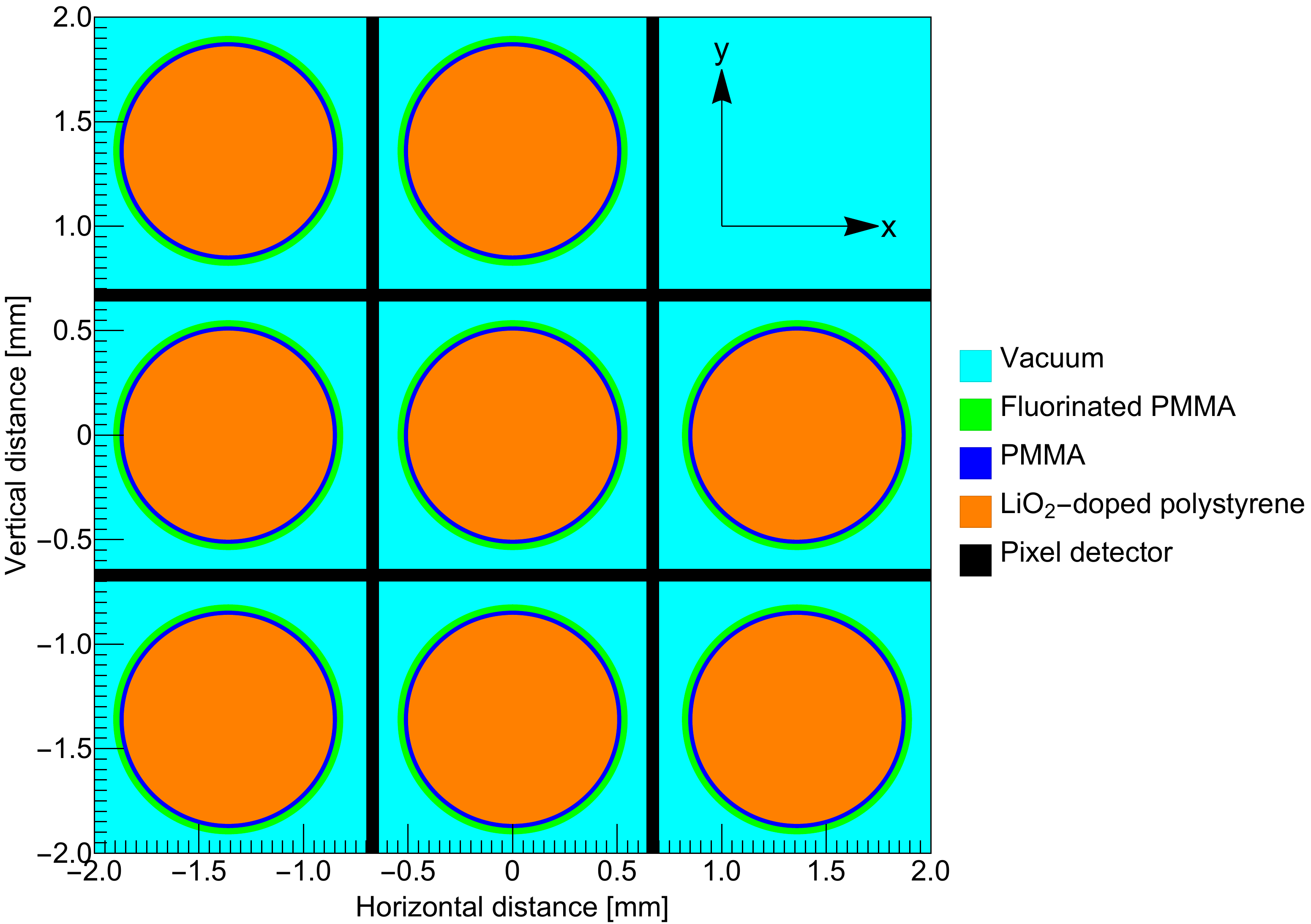}
\caption{Geometry of 10 wt.$\%$ LiO$_{2}$-doped scintillation fiber along with readout pixel in GEANT4 simulations.}
\label{Fiber_geometry}
\end{center}
\end{figure}

In the present study, we initially consider an array of the Kuraray Y11-200(M) fibers, and every array consists of 34 fibers that are horizontally placed. We change the composition of the polystyrene fiber core by adding 10 wt.$\%$ LiO$_{2}$ in terms of mass concentration~\cite{Shalom_EO}. Each of these scintillation fibers is assumed to be a double-cladded co-centered cylindrical medium where the fiber core constituting the innermost layer is manufactured from the 10 wt.$\%$ LiO$_{2}$-doped polystyrene, while the latter and last layers are the cladding materials, the chemical compositions of which are poly(methyl methacrylate) (PMMA) and fluorinated PMMA, respectively. The material densities of the layers, starting from the 10 wt.$\%$ LiO$_{2}$-doped polystyrene, are 1.15, 1.19, and 1.48 g/cm$^{3}$, respectively. The radius of the fiber core is 0.5 mm, while the radius of the first clad is 0.52 mm, and the radius of the second clad is 0.55 mm as described in Fig.~\ref{Fiber_geometry}. The length of each 10 wt.$\%$ LiO$_{2}$-doped fiber is around 5 cm, and the pitch size is 1.36 mm. The refractive indices for the fiber components that include the fiber core, the first clad, and the second clad are 1.59, 1.49, and 1.42, respectively. The energies of the optical photons are 1.75 and 2 eV by using an equal production rate, and the absorption length for the fiber elements is set as 1 m in order to increase the computation speed. Moreover, the scintillation yield is 10 optical photons per keV, and the scintillation time constant is 10 ns along with a resolution scale of 1. Also, we utilize a Birks' constant of 0.126 mm per MeV for the 10 wt.$\%$ LiO$_{2}$-doped Kuraray Y11-200(M) fiber core~\cite{heil2022new}. The surrounding medium in the current GEANT4 simulations is vacuum. All materials are defined in accordance with the GEANT4/NIST database. We use a planar vertical beam of 0.2$\times$0.2 cm$^{2}$ along the y-direction, and the incident beam consists of 10$^{5}$ neutrons by utilizing a constant kinetic energy of 0.025 eV in the case of the thermal neutrons, while the kinetic energy of the incident epithermal neutrons is 1 eV. In addition to FTFP$\_$BERT$\_$HP, EmStandardPhysics$\_$option4 is added for the generation of the optical photons~\cite{lambert2024simulation}. The secondary particle tracking inside the 10 wt.$\%$ LiO$_{2}$-doped fiber is maintained by G4Step, and the generated alpha-triton pairs are post-processed through Python. In the simulations of the optical photons, for the sake of the demonstration, the number of incident neutrons is 50. For the purpose of collecting the optical photons, we introduce a pixel grid where the dimension of each pixel pointing out every single 10 wt.$\%$ LiO$_{2}$-doped Kuraray Y11-200(M) scintillation fiber is 1.36$\times$1.36 mm$^{2}$ as illustrated in Fig.~\ref{Fiber_geometry}. The thickness of the pixel grid is 0.0236 mm, while the distance between the pixel grid and the fiber bundle is approximately 0.1 mm.
\section{Simulation outcomes}
\label{Outcomes}
\subsection{LiO$_{2}$-doped thermal neutron fiber bundle}
\begin{figure}[H]
\begin{center}
\includegraphics[width=10cm]{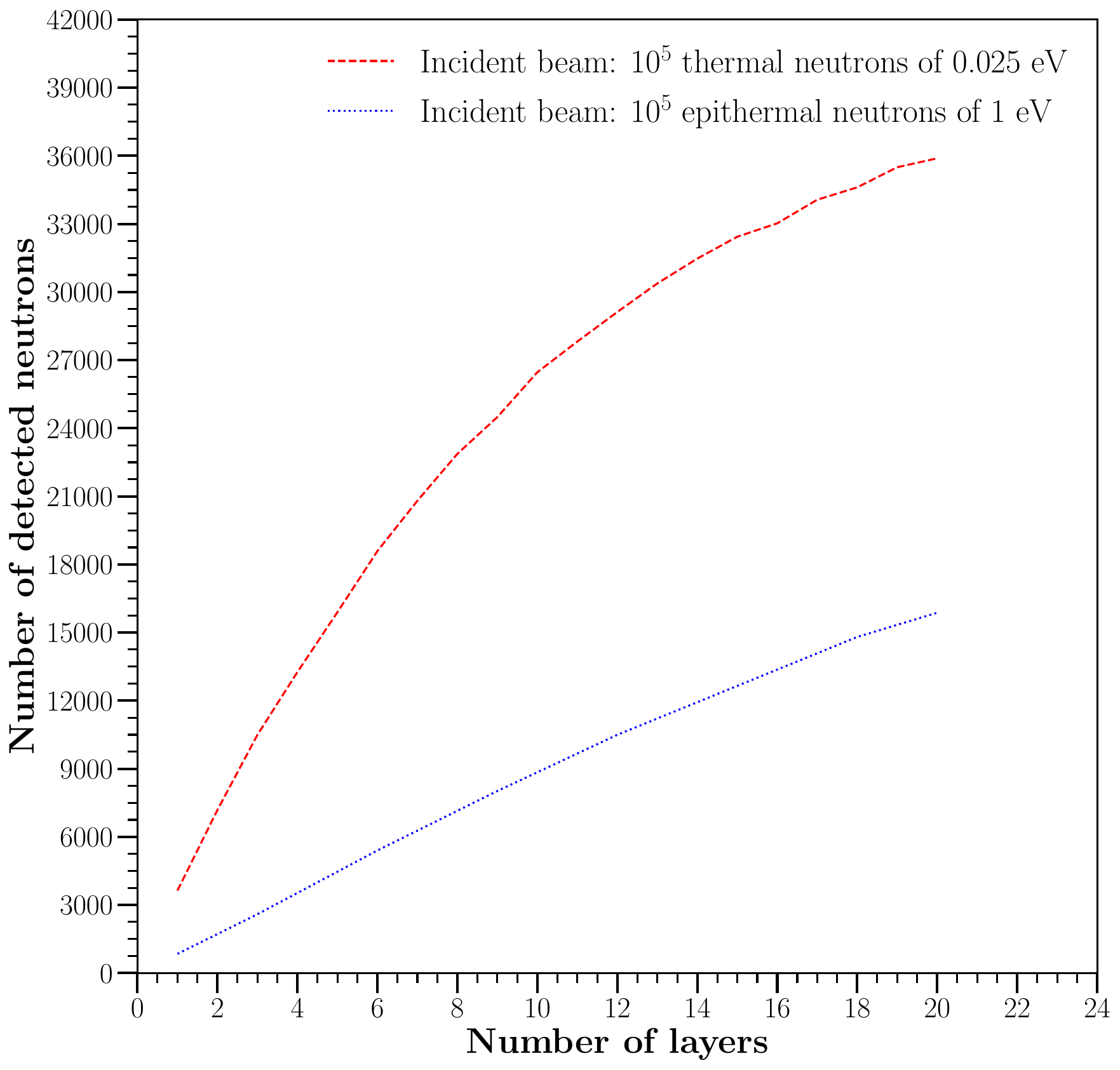}
\caption{Variation of detected neutron counts through alpha-triton pairs with respect to number of fiber layers.}
\label{Number_layers}
\end{center}
\end{figure}

\begin{figure}[H]
\begin{center}
\includegraphics[width=12cm]{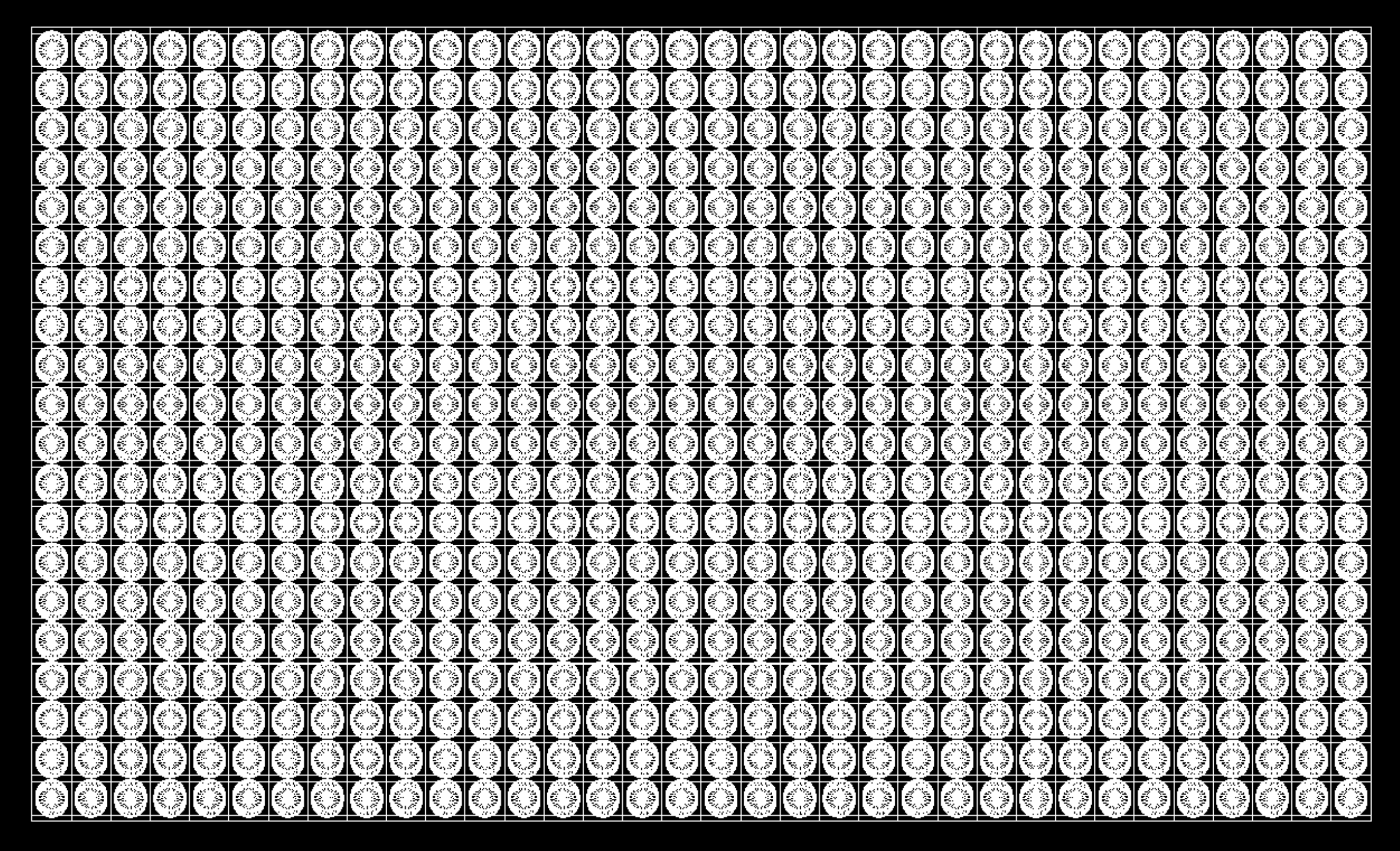}
\caption{Geometry of 20$\times$34 10 wt.$\%$ LiO$_{2}$-doped fiber bundle along with 20$\times$34 pixelated photo-sensor in GEANT4.}
\label{20_34_bundle}
\end{center}
\end{figure}

\begin{figure}[H]
\begin{center}
\includegraphics[width=8cm]{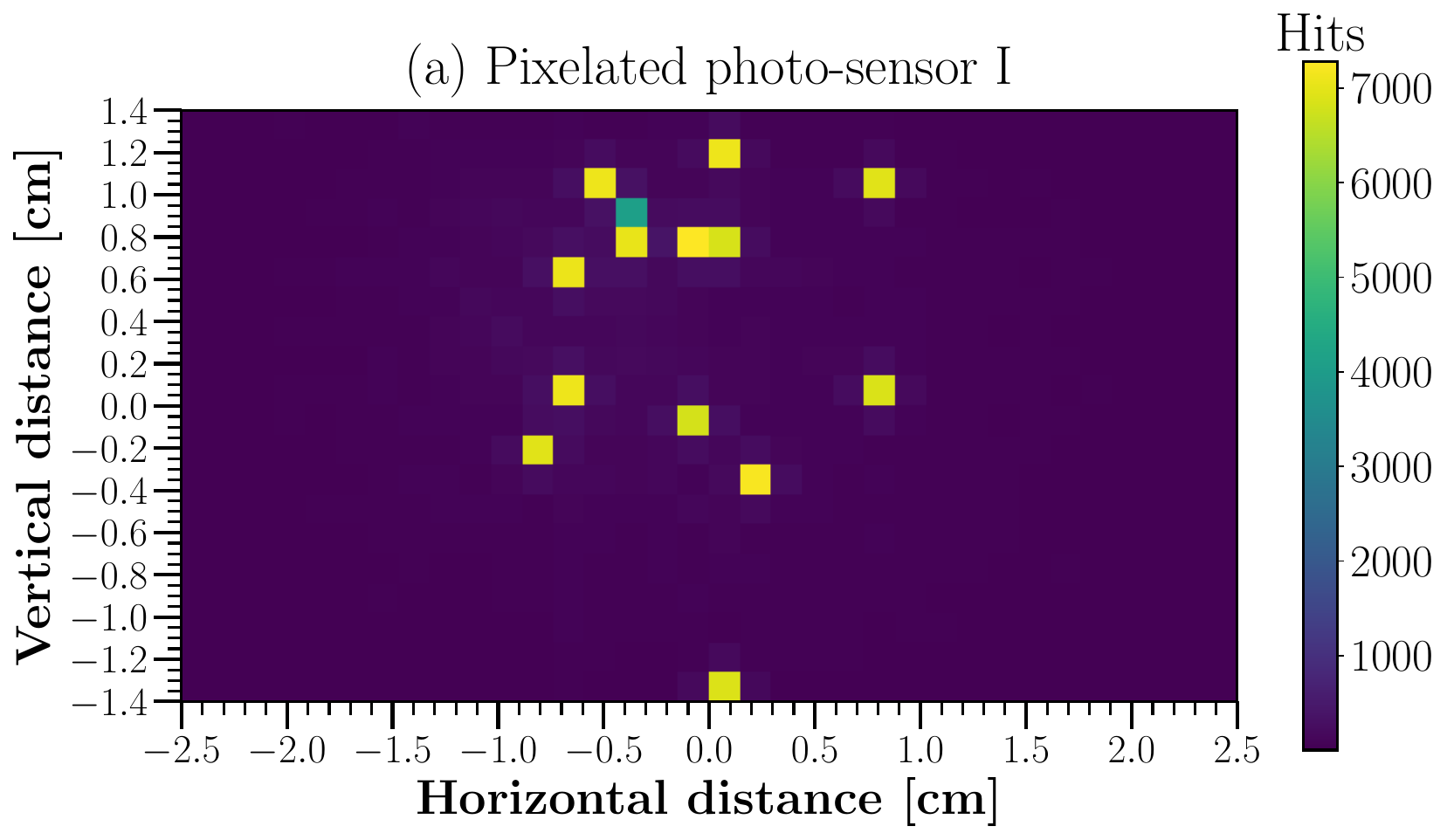}
\includegraphics[width=8cm]{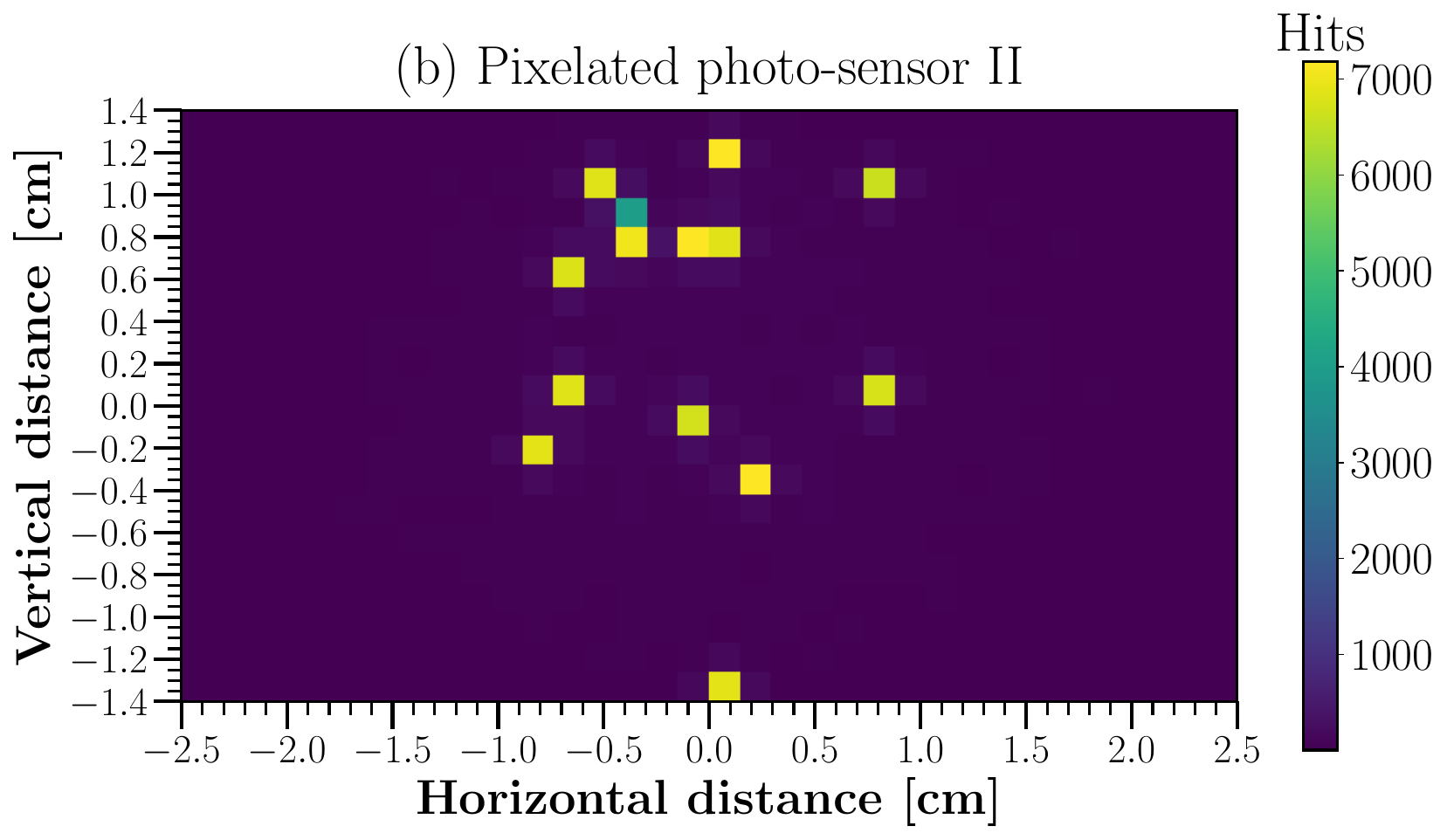}
\caption{Distribution of optical photons on 20$\times$34 pixel matrix: (a) side I and (b) side II.}
\label{Pixel_thermal}
\end{center}
\end{figure}

The fundamental principle behind the detection of the thermal/epithermal neutrons through the use of the LiO$_{2}$-doped plastic scintillators is founded on the collection of the optical photons that are emitted due to the energy deposition of the secondary alpha-triton pairs generated from the neutron capture process in the LiO$_{2}$-doped fiber core. In pursuit of assembling an optimal fiber bundle, we determine the approximate number of fiber layers beyond which the number of detected neutrons by means of the alpha-triton pairs exhibits a slower enhancement. We systematically increase the number of the fiber rows and we record the number of released alpha-triton pairs from the 10 wt.$\%$ LiO$_{2}$-doped fiber cores. As shown in Fig.~\ref{Number_layers}, in the case of the thermal neutrons with a kinetic energy of 0.025 eV, we demonstrate that the number of the detected neutrons through the generation of the alpha-triton pairs starts to show only a modest increase when the number of the fiber layers exceeds 20. Since the ratio of the detected thermal neutrons between the 20th layer and the 19th layer is 1.01, we conclude from Fig.~\ref{Number_layers} that a 10 wt.$\%$ LiO$_{2}$-doped fiber bundle consisting of 20 horizontal layers with 34 fibers in each layer is sufficient to yield a minimum detection efficiency of 30$\%$. Furthermore, considering that a continuous spectrum of neutrons in real situations might also include a significant number of epithermal neutrons, we repeat the same procedure for the epithermal neutrons of 1 eV as depicted in Fig.~\ref{Number_layers}. We observe that approximately 15$\%$ of the incident epithermal neutrons with a kinetic energy of 1 eV also generate an alpha-triton pair if the number of 10 wt.$\%$ LiO$_{2}$-doped fiber rows is 20. Thus, depending on the neutron spectrum, the presence of the epithermal neutrons brings a certain amount of uncertainty to the detection of the thermal neutrons when the present fiber bundle is employed.
In the wake of determining the optimal row number, we insert a 20$\times$34 pixelated photo-sensor at each end of the LiO$_{2}$-doped fiber bundle that collects the entrapped scintillation photons emitted owing to the energy deposition of the alpha-triton pairs as described in Fig.~\ref{20_34_bundle}. Each pixel of 1.36$\times$1.36 mm$^{2}$ on the pixelated photo-detector maps to a 10 wt.$\%$ LiO$_{2}$-doped Kuraray Y11-200(M) fiber on each side of the fiber bundle. We register the optical photons collected on every pixel and we calculate the pixel population. Furthermore, we convert the pixel population into a 20$\times$34 matrix by using the copy numbers of the pixel fibers that are assigned in GEANT4 in order to determine the optical photon distribution on the pixel grids. 
A demonstration over 50 incident thermal neutrons of 0.025 eV is shown in Fig.~\ref{Pixel_thermal}(a)-(b). From this demonstration in Fig.~\ref{Pixel_thermal}(a)-(b), we observe that 14 out of 50 thermal neutrons undergo the neutron capture process emitting an alpha-triton pair, and the total energy deposition of each alpha-triton pair, which is 4.8 MeV per alpha-triton pair, in the 10 wt.$\%$ LiO$_{2}$-doped fiber core leads to the release of approximately 7000 optical photons on each side of the 10 wt.$\%$ LiO$_{2}$-doped fiber bundle. Moreover, by recalling that the range of the alpha-triton pairs is micrometric within the 10 wt.$\%$ LiO$_{2}$-doped Kuraray Y11-200(M) scintillation fiber, almost every alpha-triton pair ceases within the fiber core, and the emitted optical photons are utterly entrapped in the corresponding fiber of emission due to the double cladding, thereby indicating that each illuminating pixel is correlated with a separate event of the neutron capture process. Hence, the pixel count, which is above a threshold of optical intensity, is a good measure of neutron hits in the fiber bundle.
\subsection{LiO$_{2}$-doped epithermal neutron fiber bundles}
\begin{figure}[H]
\begin{center}
\includegraphics[width=12cm]{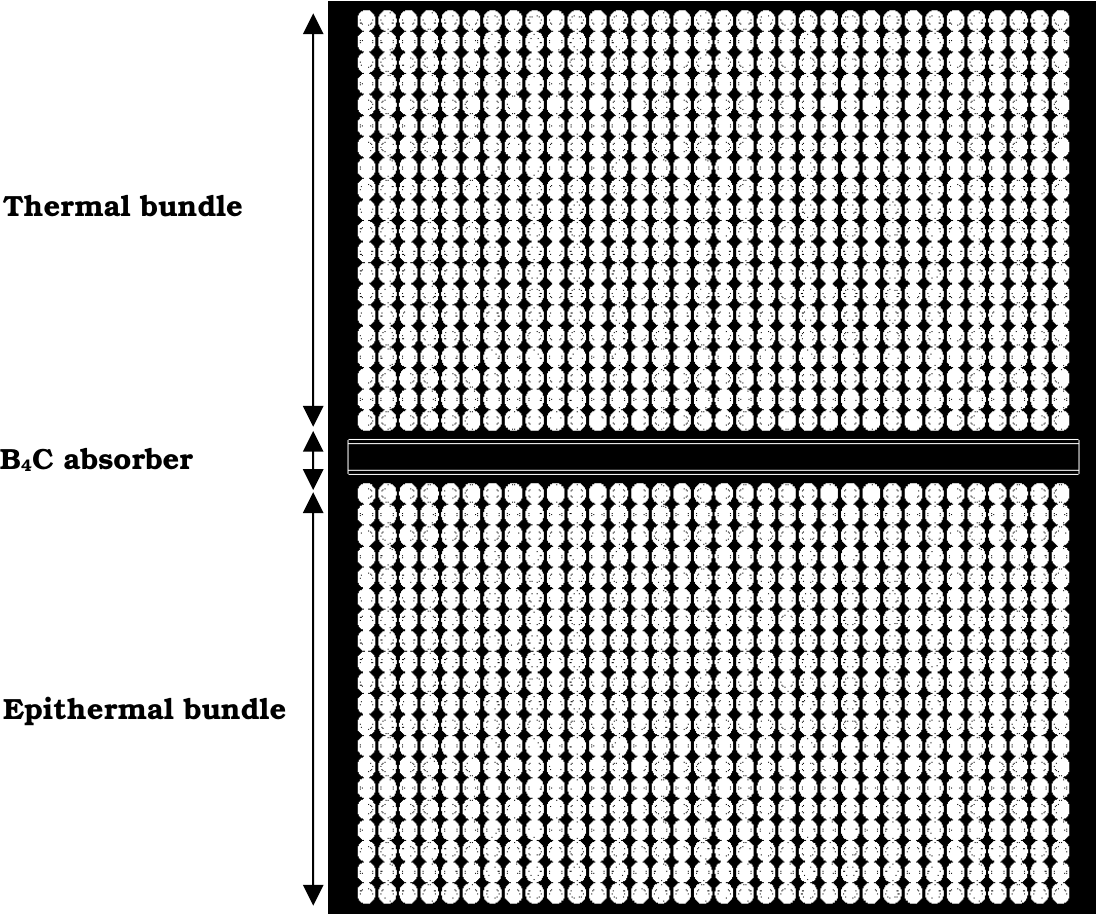}
\caption{Double 10 wt.$\%$ LiO$_{2}$-doped fiber bundles together with central 2-mm B$_{4}$C absorber layer.}
\label{Epithermal_double_bundle}
\end{center}
\end{figure}

\begin{figure}[H]
\begin{center}
\includegraphics[width=12cm]{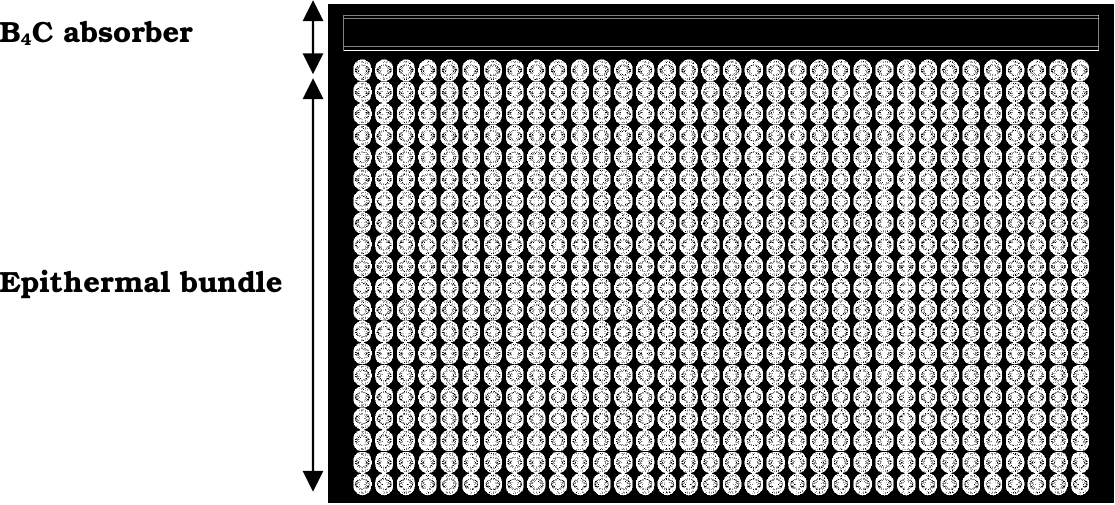}
\caption{Single 10 wt.$\%$ LiO$_{2}$-doped fiber bundle together with top 2-mm B$_{4}$C absorber layer.}
\label{Epithermal_single_bundle}
\end{center}
\end{figure}

Besides the thermal neutrons, the epithermal neutrons also induce the generation of alpha-triton pairs in the LiO$_{2}$-doped fibers since the cross-section of this process extends beyond the thermal energy levels as shown in Fig.~\ref{Number_layers}, whereby providing the opportunity to use the same LiO$_{2}$-doped fibers for the detection of epithermal neutrons. Our objective is to develop a fiber bundle system that discriminates between epithermal and thermal neutrons by minimizing the detection of thermal neutrons while maximizing the detection of epithermal neutrons. To achieve this, we first propose a combined fiber bundle system comprising a top bundle for detecting the thermal neutrons, an intermediate absorber layer to eliminate the surviving thermal neutrons after the first fiber bundle, and a bottom bundle for detecting the epithermal neutrons. This double fiber bundle includes two distinct 20$\times$34 fiber groups separated by an absorber layer that is manufactured from boron carbide (B$_{4}$C) as depicted in Fig.~\ref{Epithermal_double_bundle}. We employ a series of GEANT4 simulations to determine the optimal thickness of the B$_{4}$C absorber layer by investigating the number of detected neutrons at the bottom fiber bundle for various absorber thicknesses. We pursue an optimization criterion so as to increase the number of detected epithermal neutrons while limiting the fraction of detected thermal neutrons to approximately 5$\%$ of all thermal and epithermal detected neutrons. We use a monoenergetic planar neutron beam consisting of 10$^{5}$ neutrons in each energy regime, and our simulation outcomes for the double fiber bundle partitioned by a B$_{4}$C absorber layer are listed in Table~\ref{Tabledoublebundle}.
As demonstrated in Table~\ref{Tabledoublebundle}, a 2-mm B$_{4}$C layer effectively eliminates thermal neutrons, resulting in zero thermal neutron detection through the alpha-triton pairs at the bottom bundle while still allowing the detection of epithermal neutrons with an efficiency of approximately 0.139$\%$. Thinner absorber layers such as 1.5 and 1 mm also prevent thermal neutron detection while enhancing the epithermal neutron detection efficiency to 0.286$\%$ and 0.662$\%$, respectively. However, as the absorber layer thickness decreases further, the presence of residual thermal neutrons becomes noticeable, particularly for a 0.5-mm B$_{4}$C absorber layer, which results in 85 detected thermal neutrons alongside 1637 detected epithermal neutrons. Hence, based on our simulations, we conclude that an absorber layer thickness of 0.5 mm is optimal for securely discriminating thermal and epithermal neutrons by satisfying an ideal fraction of 0.05 in the double fiber bundle system.
\begin{table}[H]
\begin{center}
\caption{Number of detected thermal/epithermal neutrons at bottom bundle.}
\resizebox{0.9\textwidth}{!}{\begin{tabular}{*3c}
\toprule
\toprule
Thickness of B$_{4}$C [mm] & Number of detected thermal neutrons & Number of detected epithermal neutrons\\
\midrule
2.0 & 0 & 139\\
1.5 & 0 & 286 \\
1.0 & 2 & 662\\
0.5 & 85 & 1637\\
\bottomrule
\bottomrule
\label{Tabledoublebundle}
\end{tabular}}
\end{center}
\end{table}
In addition to the double fiber bundle system, we investigate an alternative configuration that involves a single LiO$_{2}$-doped fiber bundle beneath a top B$_{4}$C absorber layer for the detection of epithermal neutrons as illustrated in Fig.~\ref{Epithermal_single_bundle}. This single fiber bundle configuration aims to simplify the detection system while maintaining effective discrimination between thermal and epithermal neutrons. As shown in Fig.~\ref{Epithermal_single_bundle}, the B$_{4}$C absorber layer is positioned above the LiO$_{2}$-doped fiber bundle, and we vary its thickness to evaluate the detection efficiencies.
Our simulation results tabulated in Table~\ref{Tablesinglebundle} reveal that a top absorber layer with a thickness of 2 mm leads to the complete elimination of thermal neutron detection via the alpha-triton pairs and allows an epithermal neutron detection efficiency of approximately 0.844$\%$. Thinner absorber layers such as 1.5 and 1.0 mm also effectively remove the thermal neutrons while providing higher epithermal neutron detection efficiencies of 1.68$\%$ and 3.46$\%$, respectively. For a 0.5-mm B$_{4}$C absorber layer, the detection of thermal neutrons becomes significant, i.e. 450 thermal neutrons are detected besides 6971 detected epithermal neutrons. Consequently, similar to the double fiber bundle system, an absorber layer thickness of 0.5 mm is ideal for the single fiber bundle configuration to maximize the number of detected epithermal neutrons while keeping an optimal ratio of roughly 0.05. Finally, in accordance with the simulation outcomes in Tables~\ref{Tabledoublebundle}-\ref{Tablesinglebundle}, we demonstrate that a separate single fiber bundle combined with a B$_{4}$C is notably more effective than the double bundle configuration in the discrimination of thermal/epithermal neutrons.
\begin{table}[H]
\begin{center}
\caption{Number of detected thermal/epithermal neutrons at single bundle.}
\resizebox{0.9\textwidth}{!}{\begin{tabular}{*3c}
\toprule
\toprule
Thickness of B$_{4}$C [mm] & Number of detected thermal neutrons & Number of detected epithermal neutrons\\
\midrule
2.0 & 0 & 844\\
1.5 & 0 & 1689\\
1.0 & 7 & 3465 \\
0.5 & 450 & 6971 \\
\bottomrule
\bottomrule
\label{Tablesinglebundle}
\end{tabular}}
\end{center}
\end{table}
\section{Conclusion}
In this study, we demonstrate the feasibility of using LiO$_{2}$-doped plastic scintillating fiber bundles for detecting thermal and epithermal neutrons through the generation of secondary alpha-triton pairs. By employing the GEANT4 simulations, we determine that a 20$\times$34 fiber bundle configuration is optimal for detecting thermal neutrons with an efficiency of around 30$\%$. Additionally, we explore double and single fiber bundle systems to discriminate between thermal and epithermal neutrons. Our results indicate that a single 10 wt.$\%$ LiO$_{2}$-doped fiber bundle combined with a B$_{4}$C absorber layer is favorable by effectively eliminating the thermal neutrons while maintaining efficient epithermal neutron detection. These findings suggest that the LiO$_{2}$-doped fiber bundles can be instrumental in various applications requiring thermal/epithermal neutron detection and discrimination.
\label{Conclusion}
\bibliographystyle{elsarticle-num}
\bibliography{LiO2doped} 
\end{document}